\documentclass[10pt,conference]{IEEEtran}
\IEEEoverridecommandlockouts

\usepackage{amsfonts}
\usepackage{algorithmic}
\usepackage{graphicx}
\usepackage{textcomp}
\usepackage[]{mdframed}
\usepackage{xcolor}
\usepackage{ulem}
\usepackage{url}            
\usepackage{listings}
\usepackage{listings-golang}
\usepackage{todonotes}
\usepackage{multirow}
\usepackage{enumitem}
\usepackage{pgfplots}
\pgfplotsset{compat=1.18}
\usepackage{soul}
\def\BibTeX{{\rm B\kern-.05em{\sc i\kern-.025em b}\kern-.08em
    T\kern-.1667em\lower.7ex\hbox{E}\kern-.125emX}}

\newcommand{\ajitha}[1]{{\textcolor{teal}{}}}
\newcommand{\chao}[1]{{\textcolor{red}{}}}
\newcommand{\foivos}[1]{{\textcolor{olive}{}}}
\newcommand{\carlos}[1]{{\textcolor{blue}{}}}
\renewcommand{\baselinestretch}{0.95}
\begin{document}

\title{\texttt{Go-Oracle}: Automated Test Oracle for Go Concurrency Bugs} 

\makeatletter
\newcommand{\linebreakand}{%
  \end{@IEEEauthorhalign}
  \hfill\mbox{}\par
  \mbox{}\hfill\begin{@IEEEauthorhalign}
}
\makeatother

\author{\IEEEauthorblockN{Foivos Tsimpourlas}
\IEEEauthorblockA{
\textit{University of Edinburgh}\\
f.tsimpourlas@sms.ed.ac.uk }
\and
\IEEEauthorblockN{Chao Peng\textsuperscript{*}}
\IEEEauthorblockA{
\textit{Bytedance, China}\\
pengchao.x@bytedance.com}
\and
\IEEEauthorblockN{Carlos Rosuero}
\IEEEauthorblockA{
\textit{University of Edinburgh}\\
crosuero@ed.ac.uk}
\and
\IEEEauthorblockN{Ping Yang}
\IEEEauthorblockA{
\textit{Bytedance, China}\\
yangping.cser@bytedance.com}
\linebreakand
\IEEEauthorblockN{Ajitha Rajan\textsuperscript{*}}
\IEEEauthorblockA{
\textit{University of Edinburgh}\\
arajan@ed.ac.uk}
}

\maketitle

\begingroup\renewcommand\thefootnote{*}
\footnotetext{Corresponding authors.}
\endgroup


\begin{abstract}
The Go programming language has gained significant traction for developing software, especially in various infrastructure systems. Nonetheless, concurrency bugs have become a prevalent issue within Go, presenting a unique challenge due to the language's dual concurrency mechanisms—communicating sequential processes and shared memory. Detecting concurrency bugs and accurately classifying program executions as pass or fail presents an immense challenge, even for domain experts. We conducted a survey with expert developers at Bytedance that confirmed this challenge. 
 Our work seeks to address the test oracle problem
for Go programs, to automatically classify test executions as pass or fail. This problem has not been investigated in the literature for Go programs owing to its distinctive programming model. 

Our approach involves collecting both passing and failing execution traces from various subject Go programs. We capture a comprehensive array of execution events using the native Go execution tracer. Subsequently, we preprocess and encode these traces before training a transformer-based neural network to effectively classify the traces as either passing or failing. The evaluation of our approach encompasses 8 subject programs sourced from the GoBench repository. These subject programs are routinely used as benchmarks in an industry setting.  Encouragingly, our test oracle, \texttt{Go-Oracle}, demonstrates high accuracies even when operating with a limited dataset, showcasing the efficacy and potential of our methodology.  Developers at Bytedance strongly agreed that they would use the Go-Oracle tool over the current practice of manual inspections to classify tests for Go programs as pass or fail. 

\end{abstract}



\section{Introduction}
\label{sec:intro}

Go is a statically typed programming language designed by Google
in 2009~\cite{donovan2015go} for efficient and reliable concurrent programming. In recent years, Go has gained increasing popularity in
building software in many infrastructure systems ~\cite{grpc2018high, coredistributed, taft2020cockroachdb, ethereum2017official}. Go provides
lightweight goroutines and recommends passing messages using
channels as a less error-prone means of thread communication. However, a recent empirical study shows that concurrency bugs,
 exist widely in Go. These
bugs severely hurt the reliability of Go concurrent systems.

Test input generation techniques to help detect concurrency bugs is a promising approach~\cite{godefroid2008concurrency} but generate substantially more tests than manual approaches. This abundance poses a challenge when determining the correctness of test executions, a procedure referred to as the \emph{test oracle}, that heavily relies on manual assessment.  Surveys on the test oracle problem~\cite{barr2015oracle, nardi2015survey, langdon2017inferring} show that automated oracles based on formal specifications, metamorphic relations~\cite{liu6613484} and independent program versions are not 
widely applicable and difficult to use in practice. 
This is confirmed in the industry setting within Bytedance Research where developers use automated input generation techniques for testing Go programs that generates a large number of inputs. However, determining whether the test input execution passed or failed expectation (disregarding obvious program crashes) is still subject to manual inspection and in most cases is not trivial, taking up significant expert time. 

We seek to address the test oracle problem for Go programs that has not been investigated in the literature. 
We aim to develop \texttt{Go-Oracle}, an automated oracle that can reliably evaluate the correctness of test executions for Go programs. \texttt{Go-Oracle} is intended to act as an aid for developers when encountered with the daunting task of coming up with an expected output for each of the tests  in a large test suite. 

Designing \texttt{Go-Oracle} involves collecting a diverse dataset of both passing and failing execution traces, with labelled concurrency bug type. We use the native Go execution tracer to collect traces, that captures a comprehensive array of execution events like 
creation, start and end of Go routines,
events that block/unblock go routines (syscalls, channels, locks),
network I/O related events,
system calls and 
garbage collection. We then use these labelled traces to train a transformer model. This model effectively embeds the traces into a representation that holds essential information about the execution sequence, enabling subsequent classification as either passing or failing traces.
 It is worth noting that the paper's contribution lies in the classification of execution traces based on the presence or absence of concurrency bugs. Hence, during training, all failing traces are attributed to concurrency bugs, aligning with the focus of this research. However, the model is easily extendable to encompass classification for other bug types in the future. 


We conducted an empirical study using eight subject programs from the \texttt{GoBench} repository~\cite{yuan2021gobench} containing real-world Go concurrency bugs. 
The primary objective was to evaluate the effectiveness of \texttt{Go-Oracle} in classifying execution traces from subject programs that were not part of its training dataset. Go-Oracle demonstrated remarkable accuracy in identifying failing traces, despite the small training set, achieving a perfect accuracy of 100\% for 5 out of the 8 programs, and an impressive accuracy ranging between 86\% and 93\% for the remaining three programs.
However, in the case of identifying passing traces, the accuracy was comparatively lower, averaging at 64\%. This can be attributed to the insufficiency in labelled passing traces during the model's training phase.

We conducted a survey with three experienced Go developers from Bytedance in classifying Go tests from the eight subject programs . The survey findings revealed that the developers found manually classifying the tests as pass or fail tedious and time consuming, with varying levels of difficulty. All the developers expressed a strong preference to using the automated \texttt{Go-Oracle} tool for classifying test outcomes, despite \texttt{Go-Oracle's} lower accuracy in identifying passing traces. 

We conducted an ablation study to pinpoint the crucial components within execution traces that significantly influence classification by \texttt{Go-Oracle}. Notably, we identified that \texttt{Event} details, such as timestamp and goroutine ID within traces, played a pivotal role in classification due to their significant impact on concurrency management during program execution.

Finally, we compared performance of \texttt{Go-Oracle} against three state-of-the-art (SOTA) concurrency bug monitors that operate based on traces. We found \texttt{Go-Oracle's} accuracy in detecting failing traces significantly surpassed the performance of SOTA bug montiors. Specifically, Go-Oracle only missed 2 bugs in traces, in stark contrast to the SOTA tools that missed a substantial number of bugs, ranging from 26 to 81 out of a total of 103 bugs.

\paragraph{Source Code and Data:} We provide the source code for the Go-Oracle model, along with the subject programs and  execution traces used for training and evaluation at \\
\url{https://anonymous.4open.science/r/GoOracle-FE8C/}

\section{Background}
\label{sec:background}
In this section, we discuss basic concepts in Go programming and the concurrency mechanism utilised by Go.

\subsection{The Go Programming Language}

Go was first released in 2009 and gained popularity among developers for its simplicity, efficiency, and concurrency capabilities. The syntax of Go is similar to that of C, making it easy for developers who are familiar with C or C++ family languages. Go also has a garbage collector, which automatically frees up memory that is no longer in use, reducing the likelihood of memory leaks. With its growing community and numerous libraries, Go has become a popular language for building web applications and network servers.

\subsection{Concurrent Programming in Go}

Go is designed for concurrent programming in earnest: it treats concurrency as part of the language instead of an afterthought. Go provides two mechanisms for concurrent programming: communicating sequential processes (CSP) and concurrency through shared memory.

CSP emphasises communicating between threads rather than sharing memory and provides concurrency mechanisms to enabled communication.

\begin{itemize}
    \item \textbf{Goroutines} are lightweight threads managed by the Go runtime and allow for concurrent execution of functions. They are extremely efficient, as many goroutines can run on a single operating system thread. Goroutines are created using the \textit{go} keyword, which spawns a new goroutine to run the function in the background while the parent function continues to execute.
    \item \textbf{Channels} are used for communication and synchronisation between goroutines. They allow for safe and efficient communication between goroutines by sending and receiving messages. Channels can be used to coordinate the execution of multiple goroutines and to share data between them. Together, goroutines and channels provide a powerful concurrency model that makes it easy to write efficient, parallel code in Go.
    \item \textbf{Select statements} allow for the management of multiple channels. A goroutine can select between multiple channels and wait for the availability of a specific channel.
\end{itemize}

Go also supports traditional shared memory accesses and provides various synchronisation primitives including Mutex (\textit{lock} and \textit{unlock}), condition variable, atomic read and write operations and a primitive to wait for multiple goroutines to finish their execution.

\begin{lstlisting}[caption=A Simple Concurrency Program in Go with a Data Race Problem, label=lst:go-routine]
package main

import (
    "fmt"
)

func main() {
    data := []int{1, 2, 3, 4, 5}
    results := make(chan int)

    for _, d := range data {
        go func() {
            results <- d * 2
        }()
    }

    for i := 0; i < len(data); i++ {
        fmt.Println(<-results)
    }
}
\end{lstlisting}

Listing~\ref{lst:go-routine} shows an example of the usage of goroutines and channels. This program creates a channel to receive results (Line 9) and then spawns five goroutines (Lines 11 and 12), each of which multiplies the loop variable d by 2 and sends the result to the results channel (Line 13). The main function then waits for all the results to be received from the channel and prints them out (Line 18).

\subsection{Concurrency Bugs in Go}

Although Go is born with native concurrency support, recent studies reveal that using CSP does not guarantee concurrent Go programs free from concurrency bugs, but actually becomes a novel root cause of Go-specific concurrency bugs~\cite{tu2019understanding}. Concurrency bugs in Go can be categorised into blocking bugs and non-blocking blocks, with deadlocks and data races as representive forms respectively.

The program shown in Listing~\ref{lst:go-routine} has a data race problem because multiple goroutines are accessing and modifying the same loop variable \lstinline{d} without proper synchronisation. This can lead to unpredictable results, such as incorrect output or even a runtime panic.

To fix this issue, we need to use synchronisation mechanisms such as mutexes or channels to ensure that only one goroutine can access the shared variable at a time. For example, we could modify the program, as shown in Listing~\ref{lst:go-routine-fixed}, to use a channel (declared on Line 13) to send each loop variable \lstinline{d} to the goroutine (Line 16).

\begin{lstlisting}[caption=Fixed Program from Listing~\ref{lst:go-routine}, label=lst:go-routine-fixed]
package main

import (
    "fmt"
)

func worker(d int, results chan<- int) {
    results <- d * 2
}

func main() {
    data := []int{1, 2, 3, 4, 5}
    results := make(chan int)

    for _, d := range data {
        go worker(d, results)
    }

    for i := 0; i < len(data); i++ {
        fmt.Println(<-results)
    }
}
\end{lstlisting}

\begin{lstlisting}[caption=Structure of a Go trace file ~\cite{dvyukov2014tracer}, label=lst:go-trace]
Trace           = "gotrace" Version {Event} .
EventProcStart  = "\x00" ProcID MachineID Timestamp .
EventProcStop   = "\x01" TimeDiff .
EventFreq       = "\x02" Frequency .
EventStack      = "\x03" StackID StackLen {PC} .
EventGomaxprocs = "\x04" TimeDiff Procs .
EventGCStart    = "\x05" TimeDiff StackID .
EventGCDone     = "\x06" TimeDiff .
EventGCScanStart= "\x07" TimeDiff .
EventGCScanDone = "\x08" TimeDiff .
EventGCSweepStart  = "\x09" TimeDiff StackID .
EventGCSweepDone= "\x0a" TimeDiff .
EventGoCreate   = "\x0b" TimeDiff GoID PC StackID .
EventGoStart    = "\x0c" TimeDiff GoID .
EventGoEnd      = "\x0d" TimeDiff .
EventGoStop     = "\x0e" TimeDiff StackID .
EventGoYield    = "\x0f" TimeDiff StackID .
EventGoPreempt  = "\x10" TimeDiff StackID .
EventGoSleep    = "\x11" TimeDiff StackID .
EventGoBlock    = "\x12" TimeDiff StackID .
EventGoBlockSend= "\x13" TimeDiff StackID .
EventGoBlockRecv= "\x14" TimeDiff StackID .
EventGoBlockSelect = "\x15" TimeDiff StackID .
EventGoBlockSync= "\x16" TimeDiff StackID .
EventGoBlockCond= "\x17" TimeDiff StackID .
EventGoBlockNet = "\x18" TimeDiff StackID .
EventGoUnblock  = "\x19" TimeDiff GoID StackID .
EventGoSysCall  = "\x1a" TimeDiff StackID .
EventGoSysExit  = "\x1b" TimeDiff GoID .
EventGoSysBlock = "\x1c" TimeDiff .
EventUser       = "\x1d" TimeDiff StackID MsgLen Msg .
EventUserStart  = "\x1e" TimeDiff StackID MsgLen Msg .
EventUserEnd    = "\x1f" TimeDiff StackID MsgLen Msg .
\end{lstlisting}

\subsection{Go execution traces}

The Go execution tracer~\cite{dvyukov2014tracer} was released with Go 1.5 to allow for detailed profiling of Go programs. When enabled, it produces a compact file that encodes relevant events in a proprietary binary format, as seen in Listing~\ref{lst:go-trace}.
\ajitha{Consider removing Listing 3, does not add much value.}

\section{Related Work}
\label{sec:related}
In this section, we discuss existing work on program analysis, testing and automated test oracle for concurrency bugs in Go programs.

\subsection{Concurrency Bug Detection}

Detecting concurrency bugs has been studied by the research and industry community for decades and representative approaches include lockset-based~\cite{savage1997eraser} and happens-before analysis~\cite{o2003hybrid}.
Static lockset based concurrency bug detection technique employs race-violation rules and checks whether locks are held correctly for all shared variable accesses. Lockset-based approaches have been successfully applied to analysing C~\cite{voung2007relay} and Java~\cite{vaziri2006associating, naik2007conditional, edelstein2003framework, abadi2006types, flanagan2000type} programs to detect data races.
Happens-before based approach records read and write accesses to shared variables by tracking synchronisation events. If there exist two accesses, with one of them being write, to the same shared variable in an undetermined order, a data race error is reported.

For Go programs, Tu et al.~\cite{tu2019understanding} present the first study on concurrency bugs in Go programs and classify them into blocking and non-blocking bugs. Based on an industry-scale study on 2,100 microservices implemented in Go, Chabbi et al.~\cite{chabbi2022study} further report that the abundant usage of concurrency primitives and the language idioms themselves actually make Go programs prone to concurrency bugs.

Existing static analysis tools~\cite{zhang2021godetector, veileborg2022detecting, liu2021automatically, dilley2021automated, dilley2022bounded, tu2019understanding, staticcheck,vet} suffer from generating false alarms and do not scale to larger real-world Go projects.
For instance, Goat~\cite{veileborg2022detecting} fails on 70\% of the evaluations on real-world Go projects and has a 30\% false positive rate.
In the open-source community, Vet~\cite{vet} and StaticCheck~\cite{staticcheck} are two representative collections of static concurrency bug detectors for Go programs based on pattern matching and are specific to pre-defined bug patterns.

\paragraph{Dynamic Go Concurrency Bug Detectors:} GFuzz~\cite{liu2022goes} uses message reordering to proactively trigger concurrency bugs via order mutation, order prioritisation and runtime detection. However, it does not support other concurrency primitives such as locks and channel operators. 
Goleak\cite{goleak} is a detection tool that focuses on the state of Go routines. For each Go routine, Goleak records a stack, which includes its state, creation function, and a full execution trace. As the program executes, Goleak gathers information about each Go routine. Based on the stack trace information for the Go routines, Goleak detects the presence of different concurrency bugs (deadlocks, blocking bugs, channel misuse, data race).  For a given program, GoAT\cite{taheri2021goat} performs dynamic tracing as its first step. It instruments the program by spawning a Go routine to monitor its runtime behavior and injecting a handler before each concurrency primitive. These recorded events forms Execution Concurrency Trace (ECT), serving as the foundation for bug analysis. Secondly, GoAT conducts offline analysis of program execution based on the ECT. 

The study performed by Chabbi et al.~\cite{chabbi2022study} also reports that for complicated industrial projects, existing dynamic race detectors suffer from scalability and flakiness problems.

\subsection{Automated Test Oracle}

The most common form of test oracle is a specified oracle~\cite{barr2014oracle}. Although being effective in identifying failures, defining and maintaining such specifications is expensive. Automated test oracle validation has been studied for fields including functional, performance, security and safety testing~\cite{pezze2014automated}. Among these, the closed work related to us is that Bowring et al.~\cite{bowring2004active} proposed an active learning approach to build a classifier of programme behaviours using a frequency profile of single events in the execution trace. Evaluation of their approach was conducted over one small programme whose specific structure was well suited to their technique. 
More recently, Almaghairbe et al.~\cite{almaghairbe2017separating} proposed an unsupervised learning technique to classify unlabelled execution traces of simple programmes. 
They used agglomerative hierarchical clustering algorithms to build an automated test oracle, assuming that passing traces are grouped into large, dense clusters and failing traces into many small clusters. They evaluated their technique on 3 programmes from the SIR repository~\cite{do2005supporting}. The proposed approach has several limitations. They only support programmes with strings as inputs and do not consider correct classification of passing traces. The accuracy achieved by the technique is not high and the fraction of outputs that need to be examined by the developer is close to half of the total tests. This technique is not applicable to Go programs as the information collected is minimal and does not contain relevant information for Go concurrency bugs. 

Tsimpourlas et al.~\cite{tsimpourlas2021supervised, tsimpourlas2022embedding} proposed a technique for classifying execution traces for C programs using an LSTM architecture where the traces capture a sequence of function calls and their arguments. This approach suffers from the limitation that trace length is limited to 100 lines and, like~\cite{almaghairbe2017separating}, cannot be used for classifying Go concurrency bugs as only the function call information is used. 



Classification over execution traces is also used to detect malware~\cite{qiang2022efficient, shukla2019rnn} as static analysis tools are affected by packaging and code obfuscation techniques. Xin et al.~\cite{xin2019identifying} apply execution trace classification to understand mobile app behaviour and identify app features. 
However, test oracles for industrial case studies—Realistic programmes with complex behaviours and concurrency and input data structures have not been previously explored.

\section{Methodology}
\label{sec:methodology}
We present \texttt{Go-Oracle}, a deep learning oracle for Go concurrency test executions.  
The \texttt{GO-Oracle} model design has the following steps: 
\begin{description}
\item[Step 1:] Instrument the PUT to gather traces when executing the test inputs.
\item[Step 2:] Preprocess the traces to prune unnecessary information.
\item[Step 3:] Encode execution trace vectors into embedding representations
\item[Step 4:] Design a Transformer-based NN that learns to classify execution traces as ``passing" or ``failing".
\end{description}
In Figure~\ref{fig:overview}, we illustrate the steps in our approach. We discuss each of the steps in the rest of this Section.

\begin{figure}[htbp]
\includegraphics[scale=0.85]{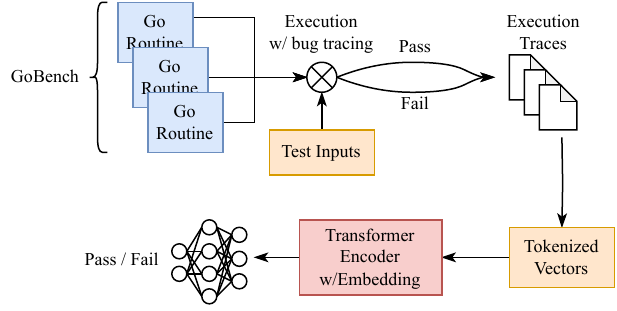}
\caption{Go-Oracle design comprising a Transformer model to summarise execution traces, trained with labelled execution traces from different Go routines, followed by a Multilayer Perceptron (MLP) to classify the trace summaries as pass or fail. }
\label{fig:overview}
\end{figure}

\subsection{Instrument and Gather Traces}
GoBench\cite{yuan2021gobench}, is a comprehensive benchmark suite for Go concurrency bugs. It comprises a total of 185 bugs, categorized into two subsets: GoReal and GoKer. The GoReal subset includes 82 real-world concurrency bugs and the GoKer subset contains 103 synthetic bugs that aim to exploit specific vulnerabilities of Go code\cite{tu2019understanding}. They exist in the form of minimal reproducible examples of the desired bug. To reproduce a GoReal bug, a Docker container clones the relevant repository, checks out the last commit known to contain the bug, and executes the test that triggers the bug. In this work, we modify GoBench to generate execution traces and label them based on the type of concurrency bug detected.

GoBench uses four tools (\textit{goleak}, \textit{go-deadlock}, \textit{dingo-hunter}, and the Go runtime's race detector \textit{Go-rd}) to check whether the bug was reproduced on a test execution. Based on the results of these checks, it classifies the test run as positive (if a concurrency bug happened) or negative (if none did). 

To maintain trace integrity without contamination from the testing environment (e.g., GoBench internal function calls), we carefully manage trace collection during the execution of the test binary. Listings \ref{lst:trace-goker} and \ref{lst:trace-goreal} illustrate the precise points in the code where trace collection was incorporated. Additionally, two utility functions, namely \lstinline{PathToTrace} (utilized in line 15 of Listing~\ref{lst:trace-goker} and line 7 of Listing~\ref{lst:trace-goreal}) and \lstinline{Trason} (utilized in line 20 of Listing~\ref{lst:trace-goker} and line 14 of Listing~\ref{lst:trace-goreal}), play key roles. \lstinline{PathToTrace} ensures that trace files are organized in a directory tree based on expected bug characteristics and test outcomes. On the other hand, \lstinline{Trason} harnesses Go's internal \lstinline{trace} package to parse trace files into JSON format for further processing. 

We end up with 80 passing and 123 failing execution traces across 8 different subject programs in our dataset. 
To ensure robustness, we train our model using an 8-fold validation spanning these subject programs. During each iteration, the model's weights are randomly initialized, and one subject program is withheld from the training data to evaluate the model's generalization accuracy.

\begin{lstlisting}[caption=Modified gobench's SingleRunResult for GoKer bugs. \\ Our contribution highlighted in yellow., label=lst:trace-goker]
func (g *GoKerExecuter) Run() *SingleRunResult {
	path := os.Getwd()
    dir = filepath.Join(filepath.Dir(path),
                        "results", "tmp")
	tmpFile := os.CreateTemp(dir, g.Bug.ID)
	tmpFile.Close()
	command := "%v -test.v -test.count %v " + 
               "-test.failfast -test.timeout %v " + 
               (*@\hl{"-test.trace \%s"}@*)
	vals := []interface{}{g.Binary, g.Count, g.Timeout,
                          (*@\hl{tmpFile.Name()}@*)}
	args := strings.Split(fmt.Sprintf(command, vals...),
                          " ")

	(*@\hl{pathToTrace := utils.PathToTrace(g.Bug.Type.String(),}@*)
                (*@\hl{g.Bug.SubType,}@*)
                (*@\hl{g.Bug.SubSubType, g.Bug.ID,}@*)
                (*@\hl{result.PositiveCheckFunc(result))}@*)
	(*@\hl{os.Rename(tmpFile.Name(), pathToTrace)}@*)
	(*@\hl{trason.Trason(pathToTrace)}@*)
	return result
}
\end{lstlisting}

\begin{lstlisting}[caption=Modified GoBench's SingleRunResult for GoReal bugs. \\ Our contribution highlighted in yellow., label=lst:trace-goreal]
func (g *GoRealExecuter) Run() *SingleRunResult {
	result := g.next()
	result.Command = fmt.Sprintf("docker exec %s %s",
                                 g.cntrCtx.Name,
                                 strings.Join(g.ExecCmd, " "))

	(*@\hl{pathToTrace := utils.PathToTrace(g.Bug.Type.String(),}@*)
                (*@\hl{g.Bug.SubType,}@*)
                (*@\hl{g.Bug.SubSubType, g.Bug.ID,}@*)
                (*@\hl{result.PositiveCheckFunc(result))}@*)

    (*@\hl{exec.Command("docker", "cp", g.cntrCtx.Name+":/tmp/trace.out",}@*)
               (*@\hl{pathToTrace).Run()}@*)
    (*@\hl{trason.Trason(pathToTrace)}@*)

	return result
} 
\end{lstlisting}


\subsection{Preprocessing}\label{preprocessing}
Every execution trace we collect is parsed into JSON format using Go's internal parser, which extracts two arrays from the execution trace file. The structure of these arrays is shown in Listing \ref{lst:trace-parse}. The \lstinline{Event}s array, encapsulates information about stack status, the program counter, the called function arguments, return values and data types and other metadata related to the event. 
The \lstinline{Stacks} array is an unordered collection of \lstinline{Frames}, each of which represents the memory stack state at the time of its associated \lstinline{Event}. 

\begin{lstlisting}[caption=ParseResult interface. Descriptions improved for clarity., label=lst:trace-parse]
type ParseResult struct {
	Events []*Event // Events is the sorted list of Events in the trace.
	Stacks map[uint64][]*Frame // Stacks is the stack traces keyed by stack IDs.
}

type Event struct {
	Off   int       // offset in input file (for debugging and error reporting)
	Type  byte      // one of Ev*
	seq   int64     // sequence number
	Ts    int64     // timestamp in nanoseconds
	P     int       // logical processor on which the event happened
	G     uint64    // goroutine on which the event happened
	StkID uint64    // unique stack ID
	Stk   []*Frame  // stack trace (can be empty)
	Args  [3]uint64 // event-type-specific arguments
	SArgs []string  // event-type-specific string args
	Link *Event     // linked event (can be nil), depends on event type
 }

type Frame struct {
	PC   uint64
	Fn   string
	File string
	Line int
}
\end{lstlisting}

\subsection{Neural Network Architecture}

We use deep neural networks (DNN) to encode our runtime execution traces and classify them as pass or fail. First, we tokenize traces into numerical vectors. We allocate one token per string keyword in the execution trace's dictionary. For numerical values, we apply a digit by digit tokenization. We prefer this method over dedicating one token per distinct numerical value because they can be very high leading to a prohibitively large and sparse vocabulary.

Next, we use an embedding layer followed by a transformer encoder architecture~\cite{DBLP:journals/corr/VaswaniSPUJGKP17} to extract features from our encoded trace information. We select the transformer as the most appropriate deep learning architecture for this task for two reasons. First, attention-based architectures are the state of the art for sequence encoding tasks, such as our execution traces that are represented as a sequence of data types and fields. Second, accuracy with a transformer model scales better as the training volume increases compared to other sequential models such as the LSTM. The transformer's only limitation is its fixed sequence length limit which we set to 4,096 tokens. Execution traces of smaller length are padded to the sequence length. Those that exceed it are truncated. For the transformer encoder, we use 2 layers with 8 attention heads and an embedding dimension size of 256.

The transformer model processes execution traces and converts them into meaningful representations within a two-dimensional feature space. This encoding process is crucial for subsequent classification.
Following the encoding step, a sequence of fully connected layers is used to reduce dimensionality and synthesize the encoded information into two key output elements. 
These output vector elements indicate the likelihood of the trace being classified as either passing or failing.

\section{Experiments}
\label{sec:experiments}
In our evaluation, we use eight repositories from GoBench, treating each repository as a distinct subject program. 
During training, we optimize model parameters tailored for classifying traces across seven of the eight subject programs. Subsequently, we evaluate the model's performance on unseen execution traces from the remaining, eighth program. We train and evaluate leaving out each of the eight programs.  This evaluation approach emphasizes the adaptability and robustness of the model across a spectrum of different repositories. We describe the configurations used in training GO-Oracle, along with the specific parameters employed in both the training and evaluation phases, to  ensure transparency of our approach.

Additionally, we conduct an ablation study. This study meticulously analyzes the sensitivity of the model's performance to various components and features, shedding light on the factors that significantly influence its effectiveness.
\begin{table}[]
    \centering
    \begin{tabular}{c|c|c}
     \hline
     \textbf{Project} & \textbf{Traces (Failing/Passing)} & \textbf{Description} \\
     \hline
     Kubernetes & 30/16 & Container manager \\ 
     Docker & 22/10 & Container framework \\
     Syncthing & 2/2 & File synchronisation system \\
     Serving & 4/6 & Serverless computing \\
     Istio & 5/9 & Service mesh \\
     CockroachDB & 28/12 & Distributed SQL database \\
     Etcd & 15/9 & Distributed key-value store \\
     Grpc-go & 16/4 & RPC library \\
     \hline
    \end{tabular}
    \vspace{5pt}
    \caption{Subject programs analysed, number of traces of each kind collected, and a short description}
    \label{tab:programs}
\end{table}

\subsection{Platforms}

We train \texttt{GO-Oracle} and conduct all our experiments on two 64-bit systems each having one Intel Xeon E5-2620 16-core CPU, 2x Nvidia GeForce GTX 1080 GPU and 32 Gigabytes of RAM. We use Ubuntu 18.04, PyTorch 1.9.1~\cite{paszke2019pytorch}, CUDA version 11.4, Nvidia driver version 510.47.03 and \texttt{Go} version 1.20.1. \texttt{GO-Tool} is a Transformer-based architecture with a sequence length of 2,048 tokens, an embedding size of 128 parameters, 2 encoder layers and 2 attention heads.

\subsection{Evaluation Setup}
We collect 8 subject programs from \texttt{GoBench} repository. Table~\ref{tab:programs} provides an overview of the subject programs, number of passing and failing traces associated with them, along with a brief description.  We instrument each repository and executed its included test cases to collect a total of 203 execution traces. We train \texttt{GO-Tool} for 2,500 steps using a batch size of 8 for each excluded subject program separately. We evaluate each of the trained model instances (with 7 of the 8 subject programs) on the execution traces that belong to the remaining unseen subject program. We use two metrics to measure the model's performance in classifying GO execution traces as passing or failing: \emph{True Negative Rate (TNR)} that measures accuracy over the negative class\footnote{passing traces are the negative class as the focus of testing is in identifying failing executions, which are the positive class} and \emph{True Positive Rate (TPR)} which indicates the model's accuracy over the positive class (failing traces). 

\ajitha{Survey Description.}
To evaluate the usefulness of Go-Oracle, we presented all the tests across 8 subjects programs and asked three developers to score the level of difficulty in manually inspecting and classifying each of the tests as pass or fail. Developer experience and their expertise is summarised in Table~\ref{tab:dev-survey}. After manual inspection, we presented Go-Oracle to the developers and the accuracy achieved in classifying the tests. We asked the developers if they would use Go-oracle in place of current practice, which is manual classification in Bytedance.
\begin{table}[]
    \centering
    \begin{tabular}{c|c|c}
     \hline
     \textbf{Developer} & \textbf{Go Experience} & \textbf{Expertise} \\
     \hline
     \texttt{Dev\#1} & 4 years & Container development \\ 
     \texttt{Dev\#2} & 3 years & Server development \\
     \texttt{Dev\#3} & 3 years & Server development \\
     \hline
    \end{tabular}
    \caption{Information on the three developers who participated in the survey for Go-Oracle evaluation}
    \label{tab:dev-survey}
\end{table}

\subsection{Comparison against State-of-the-art Tools}
Given the absence of other automated test oracles for classifying Go execution traces, our evaluation focuses on comparing \texttt{Go-oracle}'s capability to detect failing traces against SOTA dynamic Go concurrency bug detector tools, namely Goleak, GFuzz and GoAT. 
 We run each of the three tools on the 103 bugs from the GoKer dataset within GoBench. We restrict comparison to the GoKer dataset as it has smaller programs (after removing irrelevant code from GoReal) making it feasible to run all the bug detector tools to complete detection as these tools are accompanied by a high overhead. To ensure a robust evaluation, we set the frequency of executions to 100 for Goleak and GoAT and run GFuzz on the entire dataset for 10 hours.  We categorize and analyse the detection results for both blocking and non-blocking bugs. Additionally, we classify the detected bugs based on their root causes, providing a more comprehensive understanding of the tools' performance for different bug categories.

\section{Results}
\label{sec:results}
We took precautions to avoid data contamination during model training. Each model was trained separately, with all the traces from the subject program
used in testing deliberately excluded from the training set.
The evaluation results for each subject program is illustrated in Figure~\ref{fig:cross-plot}, showing the accuracy in classifying passing traces (TNR) and failing traces (TPR), as well as the total accuracy (identifying both passing and failing traces correctly).

Across subject programs, the model consistently demonstrates higher accuracy in detecting failing traces (86 -- 100\%), underlining its efficacy in identifying bugs in the majority of evaluated programs. \texttt{Go-Oracle} achieves 100\% TPR in classifying failing traces for 5 out of the 8 subject programs. The remaining 3 programs have a small number of misclassifications -- a total
of 8 misidentified failing traces as passing across the 3 programs, with TPR in the range of 86 -- 93\%. The eight misclassified traces are analysed in Section~\ref{sec:failing}.

\texttt{Go-Oracle} demonstrates varied accuracy in classifying passing traces (True Negative Rate - TNR), ranging from 50\% to 100\%. Notably, \texttt{CockroachDB} exhibits a low TNR of 33\%. Conversely, programs such as \texttt{Serving} and \texttt{Kubernetes} showcase a strong TNR when analyzed using \texttt{Go-Oracle}. The observed discrepancies in TNR across subject programs can be attributed to the imbalance in passing and failing traces available during the model's training. When subject programs have a balanced representation of passing and failing traces, \texttt{Go-Oracle} performs reasonably well. However, in cases like \texttt{CockroachDB}, where the number of passing traces is less than half the number of failing traces, \texttt{Go-Oracle}'s TNR performance is notably deficient.
Addressing this challenge could involve incorporating more accurately labeled passing traces into the training dataset. Alternatively, future work could explore techniques such as class weighting in the loss function to mitigate the impact of this imbalance, potentially enhancing \texttt{Go-Oracle}'s performance in such scenarios.
Overall, the total accuracy, measured as the fraction of passing and failing traces correctly classified by \texttt{Go-Oracle}, spans from 75\% to 100\%. Considering the limited availability of training data containing accurately labeled passing and failing Go traces, we find \texttt{Go-Oracle}'s performance to be promising. There is potential for performance improvement, especially in classifying passing traces, with a more extensive and diverse dataset.
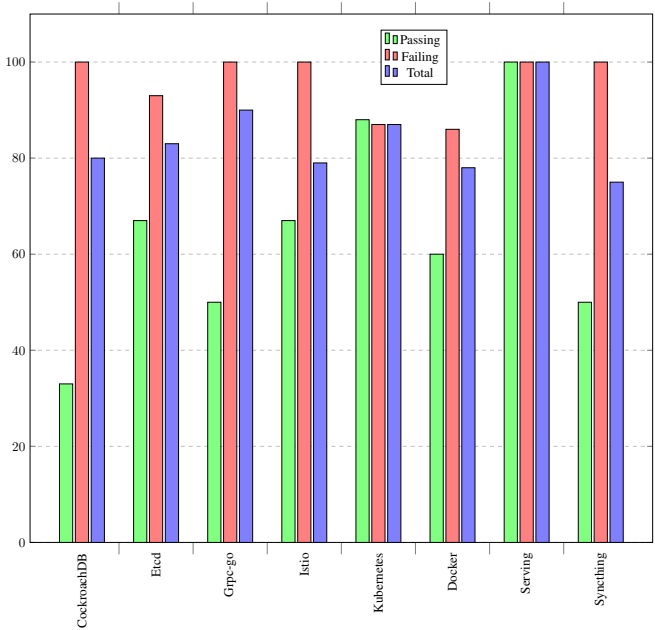
\begin{figure}
    \centering
    \pgfplotstableread{
    0 33 100 80
    1 67 93 83
    2 50 100 90
    3 67 100 79
    4 88 87 87
    5 60 86 78
    6 100 100 100
    7 50 100 75
    }\cross
    \begin{tikzpicture}[scale = 0.5]
        \begin{axis}[
            ybar,
            width=\textwidth,
            ymin=0,
            xtick=data,
            xticklabels={CockroachDB, Etcd, Grpc-go, Istio, Kubernetes, Docker, Serving, Syncthing},
            x tick label style={rotate=90},
            ytick={0,20,40,60,80,100},
            legend style={at={(0.67,0.97)}},
            ymajorgrids=true,
            grid style=dashed,
            major x tick style = {opacity=0},
            minor x tick num = 1,
            minor tick length=2ex,
        ]
            \addplot[draw=black,fill=green!50] table[x index=0,y index=1] \cross; 
            \addplot[draw=black,fill=red!50] table[x index=0,y index=2] \cross; 
            \addplot[draw=black,fill=blue!50] table[x index=0,y index=3] \cross; 
            \legend{Passing,Failing,Total}
        \end{axis}
    \end{tikzpicture}
    \caption{Labelling accuracy data for each subject program}
    \label{fig:cross-plot}
\end{figure}

\subsection{Missed Failing Traces}
\label{sec:failing}
The detection of buggy traces during testing is of paramount importance. Despite \texttt{Go-Oracle} demonstrating high accuracies in classifying failing traces, it misclassified eight failing traces as passing, shown in Table \ref{tab:missed}, across all eight subject programs.   We found failing traces containing the bugs, \texttt{grpc\#3017} and \texttt{kubernetes\#81091}, are duplicated, as they are part of both the GoKer and GoReal data sets, leading to their duplicate inclusion. Furthermore, for every bug that was not detected by the oracle, other bugs of the same characteristics were detected. An example of each is included in Table~\ref{tab:missed} for reference in the \texttt{Alternative} column. The remaining four undetected bugs in the last four rows of Table~\ref{tab:missed} stem from the Go-Real dataset, 
characterized by longer traces with a lot of irrelevant information that misleads \texttt{Go-Oracle}. Addressing this challenge entails enriching
the training dataset with similar traces that helps the model learn to focus on the more important trace parts. 

\begin{table*}[]
    \centering
    \begin{tabular}{|c|c|c|c|c|}
        \hline
        GoBench Bug ID & Category & Cause & Subcause  &Alternative\\
        \hline
        grpc\#3017 & Blocking & Resource Deadlock & Double locking & \multirow{2}{*}{syncthing\#4829} \\
        grpc\#3017 & Blocking & Resource Deadlock & Double locking  & \\
        \hline
        kubernetes\#13058 & NonBlocking & Go-Specific & WaitGroup  & \multirow{2}{*}{cockroach\#4407}\\
        kubernetes\#13058 & NonBlocking & Go-Specific & WaitGroup  & \\
        \hline
        syncthing\#5795 & Blocking & Communication Deadlock & Channel  & \multirow{2}{*}{moby\#4395}\\
        istio\#17860 & Blocking & Communication Deadlock & Channel  & \\
        \hline
        istio\#16742 & NonBlocking & Traditional & Data race  & \multirow{2}{*}{kubernetes\#81148}\\
        kubernetes\#81091 & NonBlocking & Traditional & Data race  & \\
        \hline
    \end{tabular}
    \vspace{5pt}
    \caption{Missed bugs and their characteristics}
    \label{tab:missed}
\end{table*}

\subsection{Ablation study}
To gain deeper insights into how \texttt{Go-Oracle} encodes execution trace information and identifies significant trace segments related to passing or failing traces, we conduct an ablation study. Beginning with a randomly sampled dataset with 50 passing traces and 150 failing  traces, we randomly allocate 20\% of the traces for testing purposes, reserving the remaining 80\% for model training. This separation allows for a comprehensive study without data contamination.

We commence by training the model using the complete execution trace information, setting the baseline accuracy for our ablation study. The baseline model achieves 90\% accuracy in classifying both passing (9/10 traces classified correctly) and failing (27/30 traces classified correctly) traces.   We then systematically retrain \texttt{Go-Oracle}, while using the same training and test data. During this process, we deliberately
remove one section of the trace at a time to assess
its influence on model accuracy.

We expect that the elimination of trace sections crucial to correctness to exert a significant influence on the model's accuracy. Figure~\ref{fig:enter-label} illustrates this impact, with each bar on the y-axis representing the removal of a different trace section. The zero line from the x-axis indicates traces with all the information.
Our analysis reveals a consistent decline in accuracy for both failing and passing trace detection when trace sections are removed, illustrated by bars extending to the left of zero indicating a negative effect. This underscores the critical role of specific trace properties in accurate classification. Surprisingly, removing \lstinline{Off} marginally enhances failing trace detection accuracy by 3\%. However, this improvement is deemed insignificant and likely attributed to noise given its small magnitude.
Some notable results include:
\begin{itemize}
    \item Passing trace accuracy is most affected by removing \lstinline{P}, the logical processor ID. Detection accuracy for passing falls from 90\% when all trace information is used to 50\% when \lstinline{P} is removed. We believe this is because related \lstinline{Event}s happening simultaneously on the same logical processor are more likely to result in concurrency bugs. 
    
    \item The accuracy of failing trace detection in \texttt{Go-Oracle} is notably impacted by \lstinline{Ts}, representing the \texttt{Event}'s timestamp. Upon its removal, detection accuracy drops from 90\% to 77\%. Similarly, trace information using \lstinline{P} and \lstinline{G} (the \texttt{Event}'s goroutine ID) also significantly influence failure trace detection, each resulting in an accuracy reduction of 10\% when removed. This outcome aligns with expectations, as details regarding the \texttt{Event}'s timestamp and goroutine ID are pivotal for effective goroutine management and hold the potential to cause concurrency bugs when errors arise.
    \item The overall accuracy of \texttt{Go-Oracle} experiences the most substantial impact when \lstinline{P}, representing trace information related to the logical processor ID, is removed. The accuracy decreases notably from 90\% to 73\%. This effect is observed because the trace information tied to the logical processor ID (\lstinline{P}) is critical for both passing and failing trace detection.
\end{itemize}
\begin{figure}
    \centering
    \pgfplotstableread{ 
    0 -30 3 -5
    1 -20 0 -2
    2 -10 -13 -12
    3 -40 -10 -17
    4 -20 -10 -12
    5 -10 -3 -5
    6 -30 0 -7
    }\ablation
    \begin{tikzpicture}[scale = 0.5]
        \begin{axis}[
            xbar,
            width=\textwidth,
            ytick=data,
            yticklabels={\lstinline{Off}, \lstinline{Type}, \lstinline{Ts}, \lstinline{P}, \lstinline{G}, \lstinline{StkID}, \lstinline{Stk}},
            ytick=data,
            legend pos=north west,
            xmajorgrids=true,
            yminorgrids=true,
            grid style=dashed,
            major y tick style = {opacity=0},
            minor y tick num = 1,
            minor tick length=2ex,
        ]
            \addplot[draw=black,fill=green!50] table[y index=0,x index=1] \ablation; 
            \addplot[draw=black,fill=red!50] table[y index=0,x index=2] \ablation; 
            \addplot[draw=black,fill=blue!50] table[y index=0,x index=3] \ablation; 
            \legend{Passing,Failing,Total}
        \end{axis}
    \end{tikzpicture}
    \caption{Percentage Effect on accuracy, shown on the X-axis,  of removing each attribute from the traces, shown on the Y-axis.}
    \label{fig:enter-label}
\end{figure}
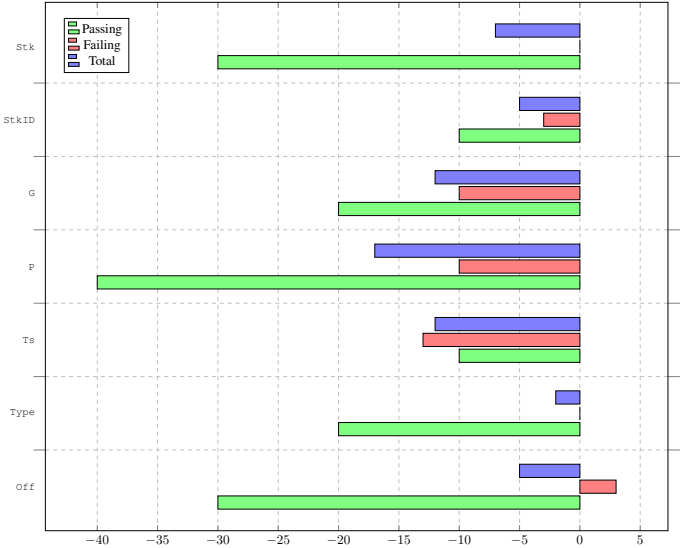

\subsection{Survey conducted with Go Developers}




\begin{figure*}[h]
\centering
\includegraphics[width=0.75\textwidth]{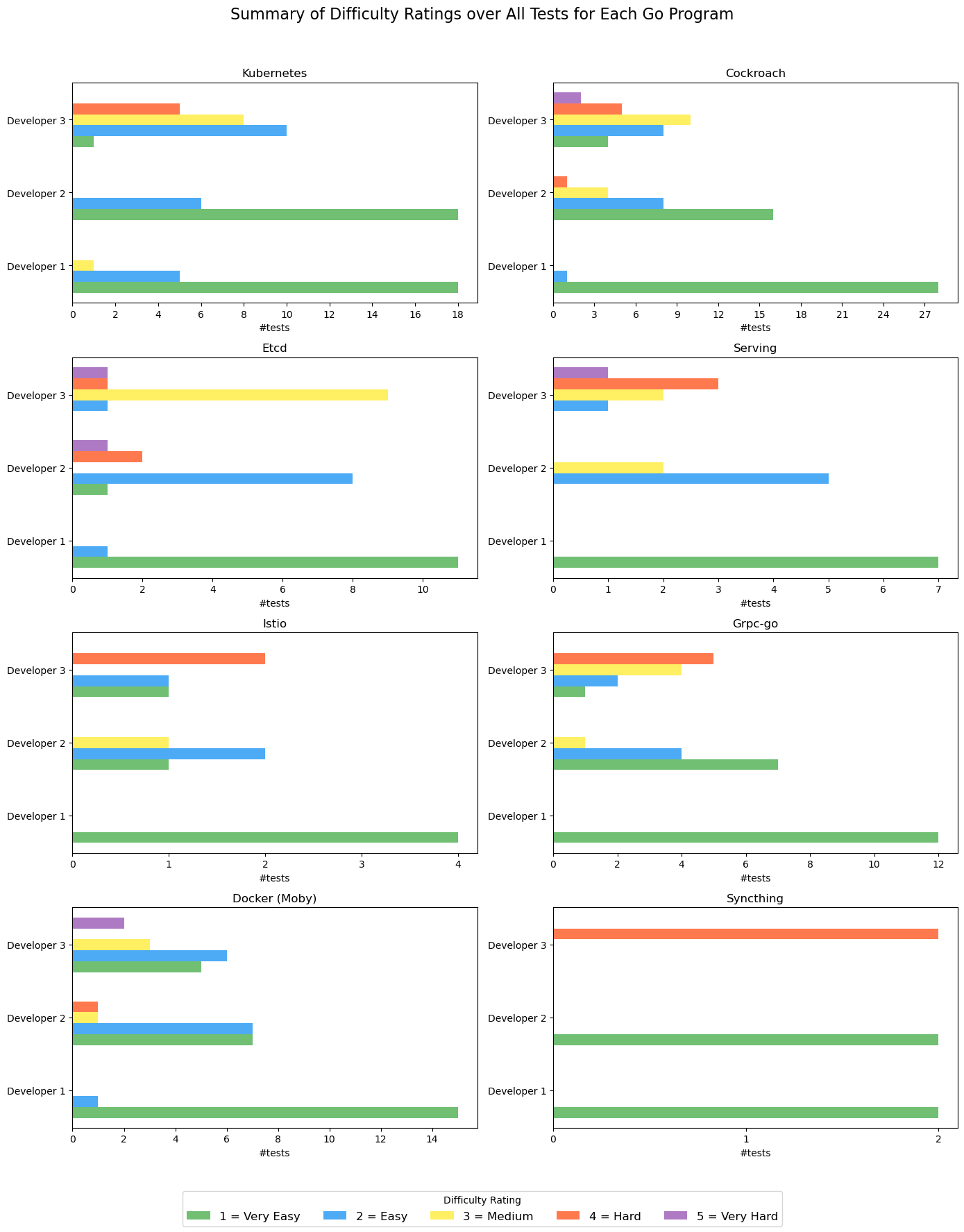}
\caption{Developer rating on difficulty of manually classifying tests as pass or fail for eight subject programs. X-axis shows numbers of tests with that difficulty rating. }
\label{fig:survey-results}
\end{figure*}

The difficulty in manually classifying tests for each of the subject Go programs as scored by the three developers is shown in Figure~\ref{fig:survey-results}. Developer 1 (\texttt{Dev\#1}) found it \texttt{very easy} to manually classify most tests as pass or fail across all subject programs. This was mainly due to his expertise in container development and familiarity with the Go subject programs that are used as a standard benchmark in industry. \texttt{Dev\#2} had slightly mixed ratings, finding tests for \texttt{Kubernetes} and \texttt{Syncthing} \texttt{easy} or \texttt{very easy} to classify. While, there is a mix of ratings for other subject programs with \texttt{easy, medium} and a few \texttt{hard} difficulty rating. \texttt{Dev\#3} found it harder to manually classify test outcomes when compared to the other two developers as seen by the mixed ratings across tests and programs in Figure~\ref{fig:survey-results}. The increased difficulty encountered by \texttt{Dev\#3} was because of lack of familiarity with the subject programs, having only worked on
the server side.  Although, the difficulty ratings varied across developers, they all concurred that the time it took to inspect and classify tests was too long. 

When asked if they would use the \texttt{Go-Oracle} tool for test classification in place of manual classification, all three developers gave the \texttt{Go-Oracle} tool highest preference for all the subject programs. The developers were impressed by the accuracy of the \texttt{Go-Oracle} tool in predicting failing test outcomes. The lower accuracy in predicting passing tests and the overhead of inspecting the false positive cases  did not cause serious concerns as the workload compared to inspecting all tests was still dramatically reduced. The developers felt certain the \texttt{Go-Oracle} tool would result in significant time savings in their routine testing process. They also expressed a strong desire for having \texttt{Go-Oracle} integrated into their workflow. The developers provided the following additional feedback (verbatim) on the current practice:
\begin{itemize}[topsep=0pt, itemsep=0pt, partopsep=0pt]
\item At ByteDance, tests similar to the survey are done in the form of end-to-end tests. Different from unit tests for functional testing that are usually a few test cases, we need to write a lot of end-to-end scenario test cases for Kubernetes-like projects.
\item Functions are created to check the final results of these tests. However, due to real-time networking issues (e.g. network speeds vary across test runs), we need to manually check the results.
\item When the network links become complicated, we need to enumerate different combinations of links and the manual inspection also becomes complicated. 
\item When a model like \texttt{Go-Oracle} is able to predict the test results, even with its current accuracy,  it will be very helpful in saving us time and we will greatly appreciate it in our workflow. 
\end{itemize}

In summary, the survey results provided evidence that there is need for an automated test oracle when testing Go programs owing to the excessive time consumed by manual inspections and classifications of test results. The survey also showed that the \texttt{Go-Oracle} tool was welcomed by developers who felt it would save significant testing time and expressed a strong preference to integrate it into their workflow.

\subsection{Comparison to SOTA Bug Detectors}
\begin{table*}[ht!]
\centering
\resizebox{0.7\textwidth}{!}{%
\begin{tabular}{ccc|ccc}
\hline
\multicolumn{3}{c|}{\multirow{2}{*}{Bug Type}}                                                                                             & \multicolumn{3}{c}{\multirow{2}{*}{Tools}}                                                         \\
\multicolumn{3}{c|}{}                                                                                                                      & \multicolumn{3}{c}{}                                                                               \\ \hline
\multicolumn{1}{c|}{\multirow{2}{*}{Category}}    & \multicolumn{1}{c|}{\multirow{2}{*}{Cause}}         & \multirow{2}{*}{Subcause(\#Num)} & \multirow{2}{*}{Goleak}  & \multirow{2}{*}{GFuzz} & \multirow{2}{*}{GoAT} \\
\multicolumn{1}{c|}{}                             & \multicolumn{1}{c|}{}                               &                                  &                                             &                        &                       \\ \hline
\multicolumn{1}{c|}{\multirow{10}{*}{Blocking}}   & \multicolumn{1}{c|}{}      & AB-BA deadlock(6)                & 2                                            & 0                      & 4                     \\
\multicolumn{1}{c|}{}                             & \multicolumn{1}{c|}{Resource}                               & Double locking(12)               & 11                                         & 0                      & 11                    \\
\multicolumn{1}{c|}{}                             & \multicolumn{1}{c|}{Deadlock}                       & RWR deadlock(5)                  & 1                                          & 0                      & 2                     \\ \cline{2-6} 
\multicolumn{1}{c|}{}                             & \multicolumn{1}{c|}{} & Channel(17)                      & 15                                         & 8                      & 13                    \\
\multicolumn{1}{c|}{}                             & \multicolumn{1}{c|}{Communication}                               & Channel \& Condition Variable(2) & 2                                             & 1                      & 0                     \\
\multicolumn{1}{c|}{}                             & \multicolumn{1}{c|}{Deadlock}      & Channel \& Context(8)            & 6                                           & 5                      & 6                     \\
\multicolumn{1}{c|}{}                             & \multicolumn{1}{c|}{}                               & Condition Variable(2)            & 2                                            & 0                      & 2                     \\ \cline{2-6} 
\multicolumn{1}{c|}{}                             & \multicolumn{1}{c|}{}         & Channel \& Lock(13)              & 7                                          & 4                      & 5                     \\
\multicolumn{1}{c|}{}                             & \multicolumn{1}{c|}{Mixed}                               & Channel \& WaitGroup(2)         & 1                                           & 0                      & 2                     \\
\multicolumn{1}{c|}{}                             & \multicolumn{1}{c|}{Deadlock}                       & Misuse WaitGroup(1)              & 1                                          & 0                      & 1                     \\ \hline
\multicolumn{1}{c|}{\multirow{6}{*}{Nonblocking}} & \multicolumn{1}{c|}{\multirow{4}{*}{Go-Specific}}   & Anonymous function(4)            & 3                                         & 0                      & 1                     \\
\multicolumn{1}{c|}{}                             & \multicolumn{1}{c|}{}                               & Misuse channel(6)                & 6                                           & 1                      & 1                     \\
\multicolumn{1}{c|}{}                             & \multicolumn{1}{c|}{}                               & Testing library(2)               & 0                                              & 0                      & 0                     \\
\multicolumn{1}{c|}{}                             & \multicolumn{1}{c|}{}                               & WaitGroup(2)                     & 1                                             & 0                      & 1                     \\ \cline{2-6} 
\multicolumn{1}{c|}{}                             & \multicolumn{1}{c|}{\multirow{2}{*}{Traditional}}   & Data race(20)                    & 18                                            & 2                      & 3                     \\
\multicolumn{1}{c|}{}                             & \multicolumn{1}{c|}{}                               & Order violation(1)                  & 1                                            & 1                      & 1                     \\ \hline
\multicolumn{1}{c|}{Overall Detection}             & \multicolumn{1}{c|}{}                               & \multicolumn{1}{c|}{Total (103)}              & 75\%                                         & 21\%                  & 51\%              \\ \hline
\end{tabular}%
}
\vspace{2pt}
\caption{Bug detection results on GoKer for the tools}
\label{tab:rq2}
\end{table*}
Table~\ref{tab:rq2} presents a summary of the detection outcomes for three SOTA tools using the bugs in the GoKer dataset. We first discuss the results from the SOTA tools for the different bug categories and then compare against \texttt{Go-Oracle}. Among the 3 dynamic tools, Goleak performs consistently better across all but three subcauses, AB-BA deadlock, RWR deadlock and Channel \& Wait Group,  where GoAT is marginally better. We analyse the three SOTA tools across the two categories, Blocking and Nonblocking bugs below. 
\paragraph{Blocking Bugs}
Goleak distinguishes itself through its comprehensive detection strategy, analyzing
the state and stack trace of each go routine. This universal approach ensures robust
detection capabilities across various bug types. Goleak proves to be a good choice when
the type of target bug is uncertain.
GoAT, as another dynamic detector, employs events analysis of different concurrency
primitives to identify bugs. Its bug detection ability across most categories is close
with those of Goleak. GoAT’s strategy allows for optimization for specific primitives,
yielding better results for primitives like WaitGroup.
GFuzz, designed exclusively for Communication Deadlock, employs message re-ordering technique to expose bugs related to message passing. Its effectiveness in this
domain is evident, with the capability to uncover a significant portion of bugs linked to
select-case statement and the enforcement of message reordering to reveal subtle bugs. However, GFuzz has limited detection ability for other
bug types.

\paragraph{Non-Blocking Bugs}
Goleak stands as the only tool equipped with detection ability
among the three tools for this bug type. It employs modules for identifying misuse of
channels and data races, exhibiting high effectiveness. However, Goleak’s performance
in uncovering nonblocking bugs arising from other causes is less satisfactory. 

\paragraph{Comparison against \texttt{Go-Oracle}}
As discussed in Section~\ref{sec:failing} and seen in Table~\ref{tab:missed}, \texttt{Go-Oracle} fails to detect a total of 8 bugs, considering both the GoReal and GoKer datasets. Notably, 2 bugs are duplicated between these datasets, resulting in a total of 6 unique missed bugs, all of which are present in the GoReal dataset. In the GoKer dataset (utilized by the state-of-the-art tools in Table~\ref{tab:rq2}), \texttt{Go-Oracle} misses only 2 bugs. One is associated with the subcause Double Locking, and the other pertains to the WaitGroup subcause. Intriguingly, all other GoKer bugs linked to failing traces are successfully detected by \texttt{Go-Oracle}. This stands in stark contrast to state-of-the-art tools, where the maximum detection rate is 75\% for Goleak, still missing 26 bugs in GoKer. Following this, GoAT detects 51\% (missing 50 bugs), and finally, GFuzz only identifies 21\% of the bugs (missing 81 bugs). In summary, \texttt{Go-Oracle} far outperforms SOTA dynamic concurrency bug detectors in accurately identifying failing traces with concurrency bugs. 


\subsection{Threats To Validity}
\label{sec:threats}
Due to the inherent challenge in detecting and labeling Go concurrency bugs, the dataset used to train our model is exceedingly limited, solely sourced from GoBench. This limitation, understandably, impacts the classification accuracy of \texttt{Go-Oracle}, contributing to the observed lower True Negative Rate (TNR) in our experiments. In the future, we will augment GoBench with an enriched dataset containing more labeled data for both passing and failing traces to mitigate this limitation.

Furthermore, with the increasing prevalence of Go programs, we anticipate that more examples of passing and failing traces will become available in open-source repositories. This broader availability of diverse data will be instrumental in enhancing the training and robustness of \texttt{Go-Oracle} for improved performance.
Another aspect impacting the validity of our results is the accuracy of the labeled data, upon which \texttt{Go-Oracle}'s accuracy is contingent. To mitigate this potential threat, we have incorporated a vetted benchmark of concurrency bugs that has been used in previous studies. This ensures that the labeled data used in training and evaluation is reliable, reducing the risk of inaccuracies influencing the outcomes of our study.

Finally, our survey was conducted with just three developers at Bytedance making it harder to generalise the results. It is, however, worth noting that the developers in the survey were experienced with Go programs in an industry setting and routinely deployed them at a large scale. Their feedback is still relevant to other industry developers. 




\section{Conclusion}
In this paper, we present \texttt{Go-Oracle}, an automated test oracle designed to classify test executions from Go routines into passing and failing traces, specifically focusing on identifying concurrency bugs. \texttt{Go-Oracle} is trained using labeled passing and failing traces from Go routines, utilizing a transformer to summarize trace information. The trace summaries are then fed into a multilayer perceptron for classification into pass or fail categories. We evaluated the effectiveness of \texttt{Go-Oracle} using eight subject programs from GoBench, containing both real and synthetic programs with concurrency bugs. Notably, \texttt{Go-Oracle} demonstrated impressive accuracy in classifying failing traces (average of 96\%), outperforming three state-of-the-art tools that monitor concurrency bugs from traces that only have a maximum bug detection accuracy of 75\%. A survey conducted with three developers at Bytedance revealed that manually classifying Go test outputs was cumbersome and time consuming, and that was the current practice followed. The developers expressed a strong preference for \texttt{Go-Oracle's} test classification and would consider integrating it into their testing pipeline. 


\vfill


\bibliographystyle{IEEEtran}
\bibliography{references}

\end{document}